\documentclass[aps,twocolumn,pra,superscriptaddress,showpacs,tightenlines]{revtex4-1}
\usepackage{amssymb}
\usepackage{amsmath}
\usepackage{graphicx}
\usepackage{subfigure}
\usepackage{amsfonts}
\usepackage{CJK}
\usepackage[colorlinks,citecolor=blue, linkcolor=blue,hyperindex,CJKbookmarks]{hyperref}

\begin{document}

\begin{CJK*}{GBK}{song}%(when sumit to arxiv, remove \begin{CJK*}{GBK}{song})

\title{Few-photon optical diode in a chiral waveguide}
\author{Jin-Lei \surname{Tan} }
\affiliation{Synergetic Innovation Center for Quantum Effects and Applications, Key Laboratory for Matter Microstructure and Function of Hunan Province, Key Laboratory of Low-Dimensional Quantum Structures and Quantum Control of Ministry of Education, School of Physics and Electronics, Hunan Normal University, Changsha 410081, China}
\author{Xun-Wei \surname{Xu}}
\affiliation{Synergetic Innovation Center for Quantum Effects and Applications, Key Laboratory for Matter Microstructure and Function of Hunan Province, Key Laboratory of Low-Dimensional Quantum Structures and Quantum Control of Ministry of Education, School of Physics and Electronics, Hunan Normal University, Changsha 410081, China}
\author{Jing \surname{Lu}}
\affiliation{Synergetic Innovation Center for Quantum Effects and Applications, Key Laboratory for Matter Microstructure and Function of Hunan Province, Key Laboratory of Low-Dimensional Quantum Structures and Quantum Control of Ministry of Education, School of Physics and Electronics, Hunan Normal University, Changsha 410081, China}
\author{Lan \surname{Zhou} }
\thanks{Corresponding author}
\email{zhoulan@hunnu.edu.cn}
\affiliation{Synergetic Innovation Center for Quantum Effects and Applications, Key Laboratory for Matter Microstructure and Function of Hunan Province, Key Laboratory of Low-Dimensional Quantum Structures and Quantum Control of Ministry of Education, School of Physics and Electronics, Hunan Normal University, Changsha 410081, China}

\begin{abstract}
We study the coherent transport of one or two photons in a one-dimensional waveguide chirally coupled
to a nonlinear resonator. Analytic solutions of the one-photon and two-photon scattering is derived.
Although the resonator acts as a non-reciprocal phase shifter, light transmission is reciprocal
at one-photon level. However, the forward and reverse transmitted probabilities for two photons
incident from either the left side or the right side of the nonlinear resonator are nonreciprocal
due to the energy redistribution of the two-photon bound state. Hence, the nonlinear resonator
acts as an optical diode at two-photon level.
\end{abstract}

%03.65.Yz 	Decoherence; open systems;
%03.65.-w 	Quantum mechanics

\maketitle

\end{CJK*}\narrowtext

\section{Introduction}
A diode that allows unidirectional propagation of signal is an indispensable circuit element
for information processing. In the design of the circuit for optical information processing,
optical isolators serving as optical diodes~\cite{JalasNPo13} are often used to realize nonreciprocal
propagation of light. Conventional optical isolators are designed based on
magneto-optic effect and work in a strong magnetic field~\cite{LBiNPo11}. The design and fabrication
of a magnetic-free optical isolator on chip have attracted a great interest in the past decades. A number of alternative schemes have been proposed theoretically and realized experimentally based on diverse mechanisms, such as nonlinear optics~\cite{LFanSci12,XZhangNPo19,RodriguezPRA19,ZWangSR15,ASZhengSR17,CaoPRL17,TudelaPRA94,LNSongOC18,PYangPRL19}, chiral quantum optics~\cite{LodahlNature17,DRoyPRA10,ScheucherSci16,HengFPRA98,ZWangPRA19}, optomechanics~\cite{ManipatruniPRL09,HafeziOE12,XWXuPRA15,MetelmannPRX15,SchmidtOPT15,LTianPRA17,ZShenNPo16,RuesinkNC16,FangNPy17,YJiangPRAPP18,DGLaiPRA20,YChenPRL21}, atomic gases in motion~\cite{DWWangPRL13,HorsleyPRL13,KXiaPRL18,SZhangNpo18,EZLiPRR20,CLiangPRL20}, and non-Hermitian optics~\cite{RuterNPy10,BenderPRL13,BPengNPy14,LChangNPo14}.
It has to be stressed that the previous studies of optical nonreciprocity have mainly focused on transmission rates of the incident classical light.
Very recently, nonreciprocal quantum effects have been explored theoretically~\cite{RHuangPRL18}, including nonreciprocal photon blockade~\cite{RHuangPRL18,BLiPRJ19,KWangPRA19,HZShenPRA20,XWXuPRJ20,XWXuPRAPP20} and nonreciprocal quantum entanglement~\cite{YFJiaoPRL20,FXSunNJP17}, which open up a way to create and manipulate one-way nonclassical light.

Here, we propose to explore the controllability of the nonreciprocal transport light in quantum regime.
Quantum switches that control the transport of a
confined single photon have been proposed in a one-dimensional (1D) waveguide~\cite{ZhouLPRA09,Lukin-np,ZLPRL08}. Strong photon-photon quantum
correlations can also be created by few-photon transport in a 1D waveguide coupled to nonlinear
scatterers, such as a two-level atom~\cite{JTShenPRL07,JTShenPRA07,SXuPRA16}, a multi-level atom~\cite{DRoyPRL11,HXZhengPRL11,HXZhengPRA12,TYLiPRA15,YPanJPA17,HXiaoACS20}, multiple two-level atoms~\cite{RephaeliPRA11,DRoyPRA13,DRoySR13,HuangPRA13}, a resonator coupling to a two-level atom~\cite{TShiPRA11,TShiPRA13,JoanesarsonPRA20}, a multi-level atom~\cite{WBYanPRA12},
or a mechanical resonator~\cite{WZJiaPRA13,LQiaoPRA17,JLiuIJTP16}. The transport
properties of a few photons inside a 1D waveguide coupled to a nonlinear resonator have been explored
theoretically, and the bunching or anti-bunching behavior as a signature of strong quantum correlations
at the few-photon level has been reported~\cite{JQLiaoPRA10,XuXWPRA2014}. However, most of the schemes are proposed based
on the configuration that a 1D waveguide coupling to a nonlinear scatterers symmetrically,
i.e., the coupling strengths are the same for photons propagating in different directions, which cannot
provide quantum non-reciprocity at the few-photon level in these systems.

In this paper, we propose a scheme to realize an optical diode at the few-photon level based on the
\emph{chiral} coupling between a 1D waveguide and a Kerr-type nonlinear resonator, i.e., the coupling
strengths between the resonator and the waveguide are different for photons propagating in different
directions. The Laplace transform is applied to solve the photon scattering problems, and an
analytic solution is obtained. It is found that, for a single photon coming from either side of the
resonator, the transmission amplitudes are the same in magnitude, but their phases are different, which
indicates that the resonator can act as a non-reciprocal phase shifter. However, the forward and reverse
transmitted probabilities for two photons incident from either the left side or the right side of the nonlinear
resonator are nonreciprocal due to the energy redistribution of the two photons bound state.
The chiral coupling configuration therefore open a new avenue toward optical diode in few-photon level
for quantum information processing on-chip.

This paper is organized as follows. In Sec.~\ref{Sec:2}, we introduce the model and
establish the notation. Then we derive the analytic solutions of the one-photon and
two-photon scattering in Sec.~\ref{Sec:3} and \ref{Sec:4}, and the forward and reverse
transmitted probabilities for photon incident from either the left side or the right
side of the nonlinear resonator are discussed at few-photon levels. We make a conclusion
in Sec.~\ref{Sec:5}.
%%%%%%%%%%%%%%%%%%%%%%%%%%%%%%%%%%%%%%%%%%%%%%%%%%%%%%%%%%%%

\section{\label{Sec:2}A chiral waveguide coupled to a nonlinear resonator}

%%%%%%%%%%%%%%%%%%%%%%%%%%%%%%%%%%%%%%%%%%%%%%%%%%%%%%%%%%%%
\begin{figure}[tbp]
\includegraphics[width=0.46\textwidth]{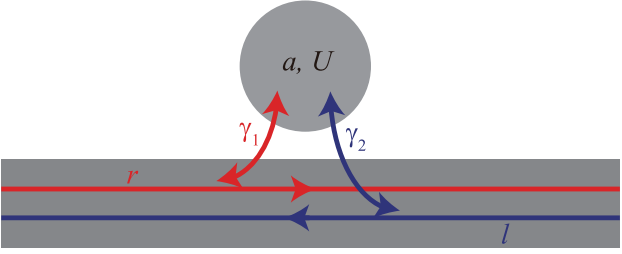}
\caption{(Color online) Schematics of a chiral 1D waveguide coupled to a nonlinear
resonator located at the origin.}
\label{Fig1.eps}
\end{figure}
The system we consider is a 1D waveguide of infinite length
coupled to a nonlinear resonator of the single-mode field at the origin, which
is schematically shown in Fig.~\ref{Fig1.eps}. The one-dimensional continuum of
the waveguide are formed by the right-going and left-going modes, the right-going
mode are represented by the canonical creation and annihilation operators $\hat{r}_{\omega }^{\dagger }$
and $\hat{r}_{\omega }$ , and the left-going mode are represented by $\hat{l}_{\omega }^{\dagger }$
and $\hat{l}_{\omega }$. Both modes obey the equal-time bosonic commutation relation
$\left[ \alpha _{\omega },\alpha_{\omega ^{\prime }}^{\dag }\right] =\delta \left( \omega -\omega ^{\prime
}\right) $ with $\alpha _{\omega }=\hat{r}_{\omega },\hat{l}_{\omega }$. Photons
traveling in the 1D waveguide can tunnel into and out of a resonator with eigenfrequency
$\omega _{c}$. The mode of the resonator is denoted by the annihilation operator $\hat{a}$.
The resonator is filled with a Kerr medium with nonlinear interaction strength $U$.
We assume that the boson velocity $v$ is frequency independent, and introduce the symbol
$\omega$ to denote the difference between the frequency of photon modes
and $\omega _{c}$. Then the Hamiltonian of the resonator-waveguide system reads
\begin{eqnarray}
\hat{H} &=&\frac{U}{2}\hat{a}^{\dagger }\hat{a}^{\dagger }\hat{a}\hat{a}+\int
\omega \left( \hat{r}_{\omega }^{\dagger }\hat{r}_{\omega }+\hat{l}_{\omega
}^{\dagger }\hat{l}_{\omega }\right) d\omega   \label{2-01} \\
&&+\int d\omega \left( \sqrt{\frac{\gamma_1 }{4\pi }}\hat{a}\hat{r}_{\omega
}^{\dagger }+\sqrt{\frac{\gamma_2 }{4\pi }}\hat{a}\hat{l}_{\omega
}^{\dagger }+h.c.\right)   \notag
\end{eqnarray}%
in the rotating frame with respective to $\hat{H}_{0}=\omega _{c}\hat{N}$, where
\begin{equation}
\hat{N}=\hat{a}^{\dagger }\hat{a}+\int_{-\infty }^{+\infty }\left( \hat{r}%
_{\omega }^{\dagger }\hat{r}_{\omega }+\hat{l}_{\omega }^{\dagger }\hat{l}%
_{\omega }\right) d\omega   \label{2-02}
\end{equation}
is called the total excitation operator and commutes with $\hat{H}$. As long as $\omega _{c}$
is far away from the cutoff frequency of the photon modes, we can extend the lower limit
of integration to $-\infty$. The first line in Eq.(\ref{2-01}) corresponds to the nonlinear
interaction with strength $U$ and free photon, the second line describes the interaction
between the waveguide and the resonator. The coupling rates to right- and left-going photons
are labeled by $\gamma_1$ and $\gamma_2$ respectively. The couplings are chiral
as long as $\gamma_1\neq\gamma_2$.

We further introduce the following operators of the waveguide
\begin{equation}
\hat{b}_{\omega }=\sqrt{\frac{\gamma_1}{\Gamma}}\hat{r}_{\omega }+\sqrt{\frac{\gamma_2}{\Gamma}}\hat{l}_{\omega }, \quad
\hat{c}_{\omega }=\sqrt{\frac{\gamma_2}{\Gamma}}\hat{r}_{\omega }-\sqrt{\frac{\gamma_1}{\Gamma}}\hat{l}_{\omega }
\label{2-03}
\end{equation}
where the total decay $\Gamma=\gamma_1+\gamma_2$. The couplings are symmetric
when $\gamma_j=\Gamma/2$. Eq.~(\ref{2-03}) decomposes the field of the waveguide into two decoupled
modes. Then, the Hamiltonian in Eq.(\ref{2-01}) is the sum of two parts $\hat{H}=\hat{H}_{c}+\hat{H}_{b}$
with
\begin{subequations}
\label{2-04}
\begin{eqnarray}
\hat{H}_{c} &=&\int_{-\infty }^{+\infty }\omega \hat{c}_{\omega }^{\dagger }\hat{c}_{\omega
}d\omega  \\
\hat{H}_{b} &=&\frac{U}{2}\hat{a}^{\dagger }\hat{a}^{\dagger }\hat{a}\hat{a}%
+\int_{-\infty }^{+\infty }\omega \hat{b}_{\omega }^{\dagger }\hat{b}%
_{\omega }d\omega  \\
&&+\sqrt{\frac{\Gamma }{\pi }}\int d\omega \left( \hat{a}\hat{b}_{\omega
}^{\dagger }+h.c.\right) d\omega   \notag
\end{eqnarray}
\end{subequations}
The $b$ modes couple to the resonator and the $c$ modes evolve freely. The
total number of excitations is conserved in both the $b$ and $c$ spaces
separately.

%%%%%%%%%%%%%%%%%%%%%%%%%%%%%%%%%%%%%%%%%%%%%%%%%%%%%%%%%%%%

\section{\label{Sec:3} scattering of  single photon}

%%%%%%%%%%%%%%%%%%%%%%%%%%%%%%%%%%%%%%%%%%%%%%%%%%%%%%%%%%%%
We study the single-photon scattering with photons incoming from either
left or right of the resonator~\cite{Lukin-np,ZLPRL08,ShenPRL05}. As the photons in the $c$ space propagate freely,
the scattering process mainly occurs in the $b$ space.

States $\hat{a}^{\dagger }\left\vert 0\right\rangle$ and $\{\hat{b}_{\omega }^{\dagger }
\left\vert 0\right\rangle\}$ are the basis in the $b$ space with one excitation,
where the ground state $\left\vert 0\right\rangle $ is a vacuum state of both
the waveguide and the resonator. $\hat{a}^{\dagger }\left\vert 0\right\rangle$
is a state for the single excitation in the resonator, $\hat{b}_{\omega }^{\dagger }\left\vert 0\right\rangle$
is a state for single excitation in the $\omega $th $b$ mode. Then, the time-dependent
wave function of the system reads
\begin{equation}
\left\vert \Psi \left( t\right) \right\rangle =A\left( t\right) \hat{a}%
^{\dagger }\left\vert 0\right\rangle +\int_{-\infty }^{+\infty }d\omega
A_{\omega }\left( t\right) \hat{b}_{\omega }^{\dagger }\left\vert 0\right\rangle
\label{3-01}
\end{equation}%
where the coefficients $A\left( t\right)$ and $A_{\omega }\left( t\right) $ are the
amplitudes of the corresponding state. The equations of motion read
\begin{subequations}
\label{3-02}
\begin{align}
i\partial _{t}A\left( t\right) & =\sqrt{\frac{\Gamma }{\pi }}\int_{-\infty
}^{+\infty }d\omega A_{\omega }\left( t\right)  \\
i\partial _{t}A_{\omega }\left( t\right) & =\omega A_{\omega }\left(
t\right) +\sqrt{\frac{\Gamma }{\pi }}A\left( t\right)
\end{align}
\end{subequations}
By perform Laplace transform, Eq.(\ref{3-02}) is transformed to a system of
algebraic equations
\begin{subequations}
\label{3-03}
\begin{align}
sA\left( s\right) -A\left( 0\right) & =-i\sqrt{\frac{\Gamma }{\pi }}\int
d\omega A_{\omega }\left( s\right)  \\
\left( s+i\omega \right) A_{\omega }\left( s\right) & =A_{\omega }\left(
0\right) -i\sqrt{\frac{\Gamma }{\pi }}A\left( s\right)
\end{align}
\end{subequations}
The scattering process indicates that initially there is no excitation in the
resonator, i.e. $A\left( 0\right) =0, A_{\omega }\left( 0\right)\neq 0$. For
the convenience of later discussion, we assume the probability amplitude at
the initial time
\begin{equation}
A_{\omega }\left( 0\right) =\frac{\sqrt{\epsilon /\pi }}{\omega -\delta
+i\epsilon }  \label{3-04}
\end{equation}
is symmetric with respect to the center $\delta $ with width $\epsilon$.
After a long time with $t>>\Gamma ^{-1},\epsilon ^{-1}$, the
resonator is in the vacuum state and the scattered phonon travels freely
in the waveguide, i.e.,
\begin{equation*}
A\left( \infty \right) =0,A_{\omega }\left( \infty \right) =\bar{t}_{\omega
}A_{\omega }\left( 0\right) e^{-i\omega t}.
\end{equation*}%
with parameter
\begin{equation}
\bar{t}_{\omega }=\frac{\omega-i\Gamma }{\omega +i\Gamma}  \label{3-05}
\end{equation}
\begin{figure}[tbp]
\includegraphics[width=8cm]{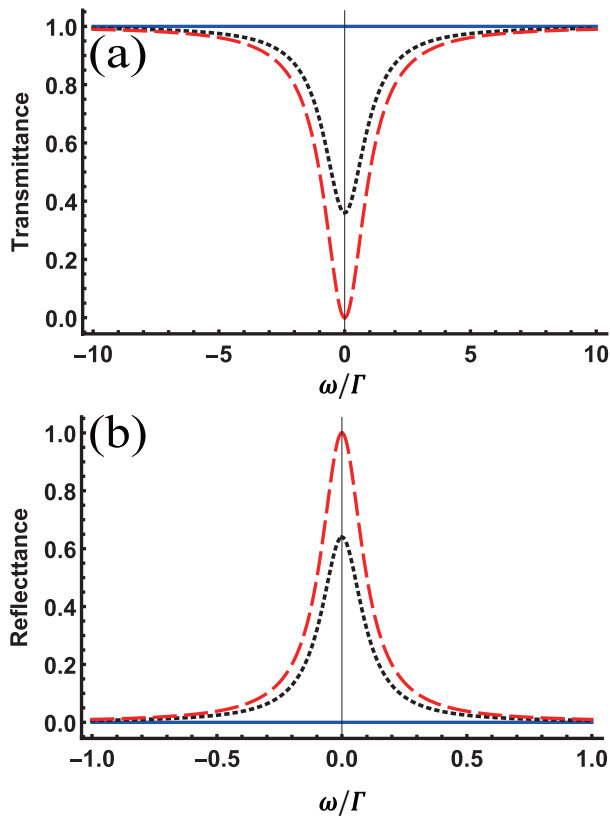}
\caption{(color online) The transmittance (a) and reflectance (b) as a function
of $\omega$. Blue solid, red dashed, and black dotted curves are shown for
$(\gamma_1,\gamma_2)=(1,0), (0.8,0.2),(0.5,0.5)$, respectively.
All parameters are in units of $\Gamma$.}
\label{fig2.eps}
\end{figure}

Let us return to the scattering process in terms of the right and left-going
modes. We first consider the single-photon wavepacket is in the right-going modes,
traveling towards the resonator at origin from the left. Its state is described by
\begin{equation*}
\left\vert \Psi \left( 0\right) \right\rangle =\int_{-\infty }^{+\infty
}d\omega A_{\omega }\left( 0\right) \hat{r}_{\omega }^{\dagger }\left\vert
0\right\rangle
\end{equation*}
Then, the photon is absorbed by the resonator. Meanwhile, the resonator emits
the photon in both the right and left directions, and relaxes to the ground state.
The single photon wavepacket will be redistributed into a reflected and a
transmitted part, which is depicted by the scattering amplitudes
\begin{equation}
\label{3-06}
t_{\omega }=\frac{\omega-i\gamma}{\omega+i\Gamma},\qquad
r_{\omega }= -2i\frac{\sqrt{\gamma_1 \gamma_2}}{\omega+i\Gamma}
\end{equation}
where we have defined $\gamma=\gamma_1-\gamma_2$. If the single-photon wavepacket
is in the left-going modes, traveling towards the resonator at origin from the right.
Its state is described by
\begin{equation*}
\left\vert \Psi \left( 0\right) \right\rangle =\int_{-\infty }^{+\infty
}d\omega A_{\omega }\left( 0\right) \hat{l}_{\omega }^{\dagger }\left\vert
0\right\rangle.
\end{equation*}
The reflection amplitude is still described by $r_{\omega }$, which can be easily understood
by the fact that the expression of $r_{\omega }$ in Eq.(\ref{3-06}) is the same under the
exchange of $\gamma_1$ and $\gamma_2$. However, the transmission amplitude becomes
\begin{equation}
\label{3-07}
t'_{\omega }=\frac{\omega+i\gamma}{\omega+i\Gamma}.
\end{equation}
Except different phases, transmission amplitudes are the same in magnitude for
an incident photon coming from either side of the resonator. For symmetric
coupling $\gamma_1=\gamma_2$, the two transmission amplitudes are equal,
i.e., $t'_{\omega }=t_{\omega }$, and we cover the results in Refs.~\cite{PRE627389,ShenOL05,ShenPRL05,ShenPRA09a,ShenPRA09b}.

Since the chiral coupling between the waveguide and the resonator introduces
an imbalance of the probabilities between the right- and left-moving emitted photon,
we plot the scattering coefficients as a function of the incident frequency $\omega$
for a fixed $\Gamma$. A Lorentzian line shape can be observed around the resonance,
however, a single photon at resonance frequency is completely reflected only for
symmetric coupling. As $\gamma$ increases, the reflection becomes lower and lower.
If the chiral coupling is fixed, the propagating single photon get nearly total
transmitted as its center is far way from the resonator mode.

%%%%%%%%%%%%%%%%%%%%%%%%%%%%%%%%%%%%%%%%%%%%%%%%%%%%%%%%%%%%

\section{\label{Sec:4} two-phonon scattering}

%%%%%%%%%%%%%%%%%%%%%%%%%%%%%%%%%%%%%%%%%%%%%%%%%%%%%%%%%%%%
We now consider the scattering process in the double-excitation subspace.
For photons incident from either side of the resonator in the $c$ space,
they get transmitted directly. For two photons incident from either side
of the resonator in the $b$ space, the two-photon wave functions have a
complicated form due to the nonlinear interaction of the resonator.
The general two-excitation state at arbitrary time in the $b$ space read
\begin{eqnarray}
\left\vert \psi (t)\right\rangle  &=&A_{c}(t)\frac{\hat{a}^{\dagger }\hat{a}%
^{\dagger }}{\sqrt{2}}\left\vert 0\right\rangle +\int d\omega B_{\omega }(t)%
\hat{a}^{\dagger }\hat{b}_{\omega }^{\dagger }\left\vert 0\right\rangle
\label{4-01} \\
&&+\int d\omega d\omega ^{\prime }C_{\omega \omega ^{\prime }}(t)\hat{b}_{\omega
}^{\dagger }\hat{b}_{\omega ^{\prime }}^{\dagger }\left\vert 0\right\rangle \theta
\left( \omega >\omega ^{\prime }\right)   \notag
\end{eqnarray}
where the function $A_{c}(t)$ stands for the amplitude of the state that both excitation
are localized in the resonator, the functions $B_{\omega }(t)$ are the amplitudes of states
with one of the excitations in the waveguide and the other in the resonator, functions
$C_{\omega \omega ^{\prime }}(t)$ are the amplitudes of states that both photons are in the
waveguide, they are symmetry to the permutation of photon frequencies $C_{\omega \omega ^{\prime }}(t)
=C_{\omega ^{\prime}\omega }(t)$ due to the bosonic nature. The initial state is expected
towards an asymptotic configuration with no excitation in the resonator and two traveling
photons in the waveguide
\begin{eqnarray}
C_{\omega \omega ^{\prime }}(0) &=&\left( \frac{C}{\omega -\delta
_{1}+i\epsilon _{1}}\frac{1}{\omega ^{\prime }-\delta _{2}+i\epsilon _{2}}%
\right.   \label{4-02} \\
&&\left. +\frac{C}{\omega ^{\prime }-\delta _{1}+i\epsilon _{1}}\frac{1}{%
\omega -\delta _{2}+i\epsilon _{2}}\right) ,  \notag
\end{eqnarray}
where the normalization constant
\begin{equation}
C=\sqrt{\frac{\epsilon _{1}\epsilon _{2}}{2\pi ^{2}}}\left[ 1+\frac{%
4\epsilon _{1}\epsilon _{2}}{\left( \delta _{1}-\delta _{2}\right)
^{2}+\left( \epsilon _{1}+\epsilon _{2}\right) ^{2}}\right] ^{-1/2}.
\label{4-03}
\end{equation}
The Schr\"{o}dinger equation yields the evolution of the amplitudes:
\begin{subequations}
\label{4-04}
\begin{align}
\partial _{t}A_{c}& =-iUA_{c}-i\sqrt{\frac{2\Gamma }{\pi }}\int d\omega B_{\omega } \\
\partial _{t}B_{\omega }& =-i\omega B_{\omega }-i\sqrt{\frac{2\Gamma }{\pi }}A_{c}-i\sqrt{\frac{\Gamma }{\pi }}\int d\omega ^{\prime }C_{\omega \omega
^{\prime }} \\
\partial _{t}C_{\omega \omega ^{\prime }}& =-i\left( \omega +\omega ^{\prime
}\right) C_{\omega \omega ^{\prime }}-i\sqrt{\frac{\Gamma }{\pi }}\left( B_{\omega}+B_{\omega ^{\prime }}\right)
\end{align}
\end{subequations}

During the scattering process the incident photons have a chance to jump
into the resonator, however, We are mainly interested in the longtime behavior
with $t\gg \gamma_i^{-1}, \epsilon_i^{-1}$. The asymptotic solution of $C_{\omega \omega ^{\prime }}$
is obtained by first Laplace transform of Eqs.(\ref{4-04}) and then via the
inverse Laplace transform up to the forth order of $\sqrt{\Gamma}$ after some
algebra calculations. It has an overall phase factor $%
\exp \left[ -i\left( \omega +\omega ^{\prime }\right) t\right] $ multiplying
an time-independent part
\begin{equation}
C_{\omega \omega ^{\prime }}(t)=e^{-i\left( \omega +\omega ^{\prime }\right) t}
\left[ \bar{t}_{\omega}\bar{t}_{\omega ^{\prime }}C_{\omega \omega ^{\prime }}(0)
+D_{\omega \omega ^{\prime }} \right]
\label{4-05}
\end{equation}
where we have introduced
\begin{eqnarray}
D_{\omega \omega ^{\prime }}&&=\frac{U}{\omega+\omega ^{\prime }-U+2i\Gamma}\frac{2C(\bar{t}_{\omega}-1)
(\bar{t}_{\omega^{\prime }}-1)}{\omega+\omega ^{\prime }-\delta_1-\delta_2+i(\epsilon_1+\epsilon_2)} \\
&&\times \left[\frac{1}{\omega+\omega ^{\prime }-\delta_1+i(\epsilon_1+\Gamma)}+ \frac{1}{\omega+\omega ^{\prime }-\delta_2+i(\epsilon_2+\Gamma)}\right]
\notag
\label{4-06}
\end{eqnarray}

For two incoming right-going photons from the left of the resonator with
the initial state%
\begin{equation}
\left\vert \psi (0)\right\rangle =\int d\omega d\omega ^{\prime }C_{\omega
\omega ^{\prime }}(0)\hat{r}_{\omega }^{\dagger }\hat{r}_{\omega ^{\prime }}^{\dagger
}\left\vert 0\right\rangle,   \label{4-07}
\end{equation}%
there are four possibilities after the two incoming photons with frequencies $\omega $
and $\omega ^{\prime }$ colliding with the resonator: Both are transmitted
with amplitude $\alpha_{\omega \omega ^{\prime }}^{rr}$; Both are reflected with
amplitude $\alpha_{\omega \omega ^{\prime }}^{ll}$; One photon is transmitted and
the other is reflected with amplitude $\alpha_{\omega \omega ^{\prime }}^{rl}$
and $\alpha_{\omega \omega ^{\prime }}^{lr}$. The state of out-going photons reads%
\begin{eqnarray*}
\left\vert \psi (t)\right\rangle  &=&\int d\omega d\omega ^{\prime }\left(
\alpha_{\omega \omega ^{\prime }}^{rr}\hat{r}_{\omega }^{\dagger }\hat{r}_{\omega ^{\prime
}}^{\dagger }+\alpha_{\omega \omega ^{\prime }}^{ll}\hat{l}_{\omega }^{\dagger
}\hat{l}_{\omega ^{\prime }}^{\dagger }\right) e^{-i\left( \omega +\omega ^{\prime
}\right) t}\left\vert 0\right\rangle  \\
&&+\int d\omega d\omega ^{\prime }\left( \alpha_{\omega \omega ^{\prime
}}^{rl}\hat{r}_{\omega }^{\dagger }\hat{l}_{\omega ^{\prime }}^{\dagger }+\alpha_{\omega
\omega ^{\prime }}^{lr}\hat{l}_{\omega }^{\dagger }\hat{r}_{\omega ^{\prime }}^{\dagger
}\right) e^{-i\left( \omega +\omega ^{\prime }\right) t}\left\vert
0\right\rangle
\end{eqnarray*}
after a long time, the amplitudes are related to the single photon scatter
amplitudes $\left\{ t_{\omega },r_{\omega }\right\} $ by
\begin{subequations}
\label{4-08}
\begin{eqnarray}
\alpha_{\omega \omega ^{\prime }}^{rr} &=&t_{\omega }t_{\omega ^{\prime
}}C_{\omega \omega ^{\prime }}(0)+\frac{\gamma_1^2}{\Gamma^2}D_{\omega \omega ^{\prime }} \\
\alpha_{\omega \omega ^{\prime }}^{ll} &=&r_{\omega }r_{\omega ^{\prime
}}C_{\omega \omega ^{\prime }}(0)+\frac{\gamma_1\gamma_2}{\Gamma^2}D_{\omega \omega ^{\prime }} \\
\alpha_{\omega \omega ^{\prime }}^{rl} &=&t_{\omega }r_{\omega ^{\prime
}}C_{\omega \omega ^{\prime }}(0)+\frac{\sqrt{\gamma_1^3\gamma_2}}{\Gamma^2}D_{\omega \omega ^{\prime }} \\
\alpha_{\omega \omega ^{\prime }}^{lr} &=&r_{\omega }t_{\omega ^{\prime
}}C_{\omega \omega ^{\prime }}(0)+\frac{\sqrt{\gamma_1^3\gamma_2}}{\Gamma^2}D_{\omega \omega ^{\prime }}
\end{eqnarray}
\end{subequations}
The first term on the right side of each line of Eq.(\ref{4-08}) is the contribution from two
noninteracting photons. The second part presents the correlation of output photons coming from
the nonlinearity in the resonator, which is due to the existence of two-photon bound state\cite{JQLiaoPRA10,XuXWPRA2014}.
\begin{figure}[tbp]
\includegraphics[width=8cm]{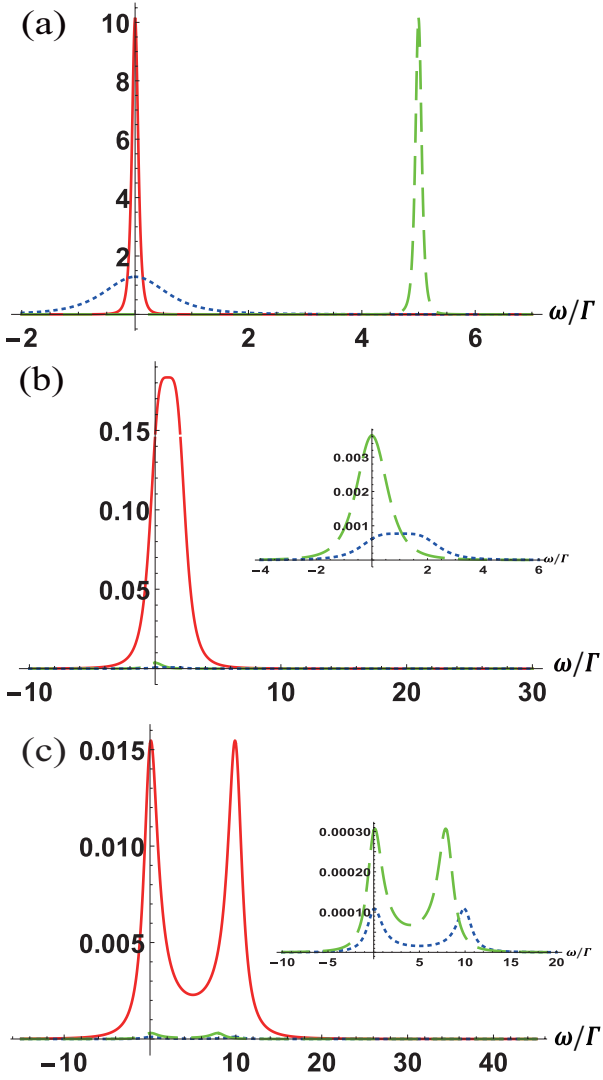}
\caption{(Color online) The probability of the initial state (red solid line and green dashed line in (a))
and the norm square of $D_{\omega \omega ^{\prime }}$ as function the one photon frequency $\omega$ for
given $U=10, \epsilon=0.1$ in (a),(c) and $U=2, \epsilon=0.1$ in (b). In (a), the red solid line and the blue dotted
line is for $(E,\delta)=(0,0)$, the green dashed line is for $(E,\delta)=(10,5)$. $(E,\delta)=(2,1)$
$(0,1)$ and $(2,0)$ for red solid line,green dashed line and blue dotted line in (b), and $(E,\delta)=(10,5)$,
$(8,5)$ and $(10,4)$ for red solid line,green dashed line and blue dotted line in (c) respectively. All
parameters are in units of $\Gamma$.}
\label{fig3.eps}
\end{figure}

For two incoming left-going photons from the right of the resonator with
the initial state
\begin{equation}
\left\vert \psi (0)\right\rangle =\int d\omega d\omega ^{\prime }C_{\omega
\omega ^{\prime }}(0)\hat{l}_{\omega }^{\dagger }\hat{l}_{\omega ^{\prime }}^{\dagger
}\left\vert 0\right\rangle,   \label{4-09}
\end{equation}
the asymptotic state of the field for a long time after two photons colliding
with the resonator can be written as
\begin{eqnarray*}
\left\vert \psi (t)\right\rangle  &=&\int d\omega d\omega ^{\prime }\left(
\beta_{\omega \omega ^{\prime }}^{rr}\hat{r}_{\omega }^{\dagger }\hat{r}_{\omega ^{\prime
}}^{\dagger }+\beta_{\omega \omega ^{\prime }}^{ll}\hat{l}_{\omega }^{\dagger
}\hat{l}_{\omega ^{\prime }}^{\dagger }\right) e^{-i\left( \omega +\omega ^{\prime
}\right) t}\left\vert 0\right\rangle  \\
&&+\int d\omega d\omega ^{\prime }\left( \beta_{\omega \omega ^{\prime
}}^{rl}\hat{r}_{\omega }^{\dagger }\hat{l}_{\omega ^{\prime }}^{\dagger }+\beta_{\omega
\omega ^{\prime }}^{lr}\hat{l}_{\omega }^{\dagger }\hat{r}_{\omega ^{\prime }}^{\dagger
}\right) e^{-i\left( \omega +\omega ^{\prime }\right) t}\left\vert
0\right\rangle
\end{eqnarray*}
The amplitudes are related to the single-photon transmission and reflection
$\left\{ t'_{\omega },r_{\omega }\right\} $ by
\begin{subequations}
\label{4-10}
\begin{eqnarray}
\beta_{\omega \omega ^{\prime }}^{ll} &=&t'_{\omega }t'_{\omega ^{\prime
}}C_{\omega \omega ^{\prime }}(0)+\frac{\gamma_2^2}{\Gamma^2}D_{\omega \omega ^{\prime }} \\
\beta_{\omega \omega ^{\prime }}^{rr} &=&r_{\omega }r_{\omega ^{\prime
}}C_{\omega \omega ^{\prime }}(0)+\frac{\gamma_1\gamma_2}{\Gamma^2}D_{\omega \omega ^{\prime }} \\
\beta_{\omega \omega ^{\prime }}^{rl} &=&t'_{\omega }r_{\omega ^{\prime
}}C_{\omega \omega ^{\prime }}(0)+\frac{\sqrt{\gamma_2^3\gamma_1}}{\Gamma^2}D_{\omega \omega ^{\prime }} \\
\beta_{\omega \omega ^{\prime }}^{lr} &=&r_{\omega }t'_{\omega ^{\prime
}}C_{\omega \omega ^{\prime }}(0)+\frac{\sqrt{\gamma_2^3\gamma_1}}{\Gamma^2}D_{\omega \omega ^{\prime }}
\end{eqnarray}
\end{subequations}
The first term of the amplitudes $\beta_{\omega \omega ^{\prime }}^{ll}$ (a transmitted pair),
$\beta_{\omega \omega ^{\prime }}^{rr}$ (a reflected pair), $\beta_{\omega \omega ^{\prime }}^{rl}$
and $\beta_{\omega \omega ^{\prime }}^{lr}$  (a pair of
one transmitted and one reflected), is also the contribution from two noninteracting photons, which is in the same
magnitude to the corresponding term for the both transmitted, both reflected, one transmitted and one
reflected amplitudes in Eq.(\ref{4-08}). However, The second part of each amplitude is different in
magnitude to the corresponding term in Eq.(\ref{4-08}) as long as $\gamma_1\neq \gamma_2$. In this sense,
the nonlinear resonator acts as an optical diode for two-photon transport.

\begin{figure}[tbp]
\includegraphics[width=8cm]{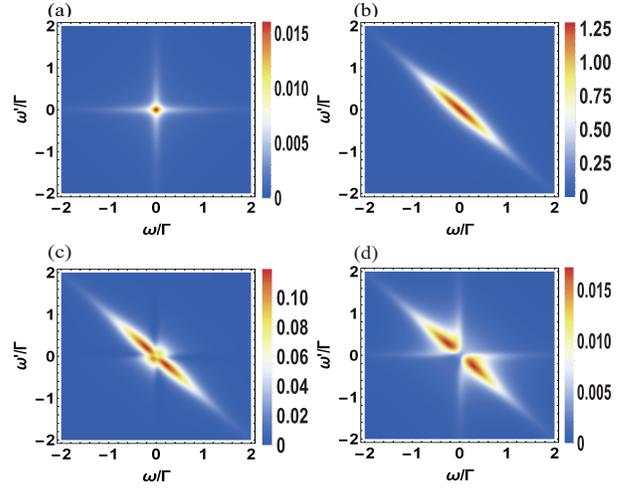}
\caption{(Color online) The norm square of the amplitudes of (a) the NP part, (b) $D_{\omega \omega'}$,
and (c) $\alpha_{\omega \omega'}^{rr}$, (d) $\beta_{\omega \omega'}^{ll}$ of the two transmitted photons as
functions of $\omega,\omega'$ with $U=10, \delta=0, \epsilon=0.1, (\gamma_1,\gamma_2)=(0.6,0.4)$. All parameters
are in units of $\Gamma$.}
\label{fig4.eps}
\end{figure}

We now assume that two photons are initially independent of each other, i.e., $\delta_1=\delta_2=\delta$ and $\epsilon_1=\epsilon_2=\epsilon$. In this case, the initial state
of the two photons becomes
\begin{equation}
C_{\omega \omega ^{\prime }}(0)=\frac{\epsilon}{\pi}\frac{1}{\omega-\delta+i\epsilon}\frac{1}{\omega^{\prime}-\delta+i\epsilon},   \label{4-11}
\end{equation}
which has its extremum at $\omega,\omega^{\prime }=\delta$. Eqs.(\ref{4-08}) and (\ref{4-10}) indicate
that the interference is produced by two waves: one wave describes the two noninteracting photons (denoted as NP part),
the other wave describes the two interacting photons (denoted as IP part) which is proportional to
\begin{eqnarray}
D_{\omega \omega ^{\prime }}&&=\frac{U}{\omega+\omega ^{\prime }-U+2i\Gamma}\frac{\epsilon/\pi}{\omega+\omega ^{\prime}-2\delta+i2\epsilon} \\
&&\times\frac{i2\Gamma}{\omega+i\Gamma}\frac{i2\Gamma}{\omega^{\prime}+i\Gamma} \frac{2}{\omega+\omega ^{\prime }-\delta+i(\epsilon+\Gamma)}.
\notag
\label{4-12}
\end{eqnarray}
Once the nonlinear interaction $U=0$, the IP part disappears. The value of the IP part are strongly
dependent on the total energy $E=\omega+\omega^{\prime}$ of the two incident photons for a given $U\neq0$.
Its maximum may occur at $E=0, \delta, 2\delta, U$ respectively. In Fig.\ref{fig3.eps}, we have plotted
the  probability of the initial state  (red solid line and green dashed line in (a)) and the norm square
of $D_{\omega \omega ^{\prime }}$((b)-(c), and blue dotted line in (a)) as function of the frequency $\omega$
for given $U, \epsilon$. It can be observed that the IP part achieves its maximum value at $\delta=0$ and
$(\omega,\omega^{\prime })=(0,0)$, i.e., both the total energy and the center frequency of each photon is
resonant with the resonator, however the incident wave also achieves its maximum value at $(\omega,\omega^{\prime })=(0,0)$.
It indicates that the interference between the NP part and the IP part occurs in the region around
$(\omega,\omega^{\prime })=(0,0)$. In Fig.\ref{fig4.eps}, we have plotted the probability of the NP
part (a), the IP part (b) of the two transmitted photons, and the norm square of  the amplitude
$\alpha_{\omega \omega'}^{rr}$ (c) and amplitude $\beta_{\omega \omega ^{\prime }}^{ll}$ (d) as functions
of $\omega,\omega'$.
The different phases of both transmission amplitude introduced by the chiral coupling produce the
different interference pattern of two transmitted photons incident from the right-going mode and
the left-going mode.

In Fig.~\ref{fig3.eps}(b-c), we have set $E=U=2\delta$ for the red solid line and $E\neq U=2\delta$ for
green dashed line, $E=U\neq2\delta$ for blue dotted line, where all $\delta\neq0$. It can be found that
the maximum value of the IP part is achieved at $E=2\delta=U$, i.e., the center frequency of each
photon is far away from the resonator mode, the sum of the centers of the two photons is equal to the
energy of the resonator containing two photons and the total energy of the incident photon satisfies the
two-photon resonance. When the nonlinear strength $U$ is in the order of magnitude of $\Gamma$, e.g.
panel (b), the IP part has its maximum value around the $(\omega,\omega')=(1,1)$ where the incident
state also achieves its maximum value, but the width of the IP part is larger than that of the incident
state. When the nonlinear strength $U$ is much larger than $\Gamma$,  e.g. panel (c), the bound state
achieves its maximum value around the $(\omega,\omega')=(0,10), (10,0)$ however, the incident wave achieves
its maximum value at $(\omega,\omega')=(5,5)$. Hence, the IP part gives rise to a redistribution of the
energy of the photons. It is known in the previous section that the propagating single photon with the
center around $\delta>\Gamma$ and width $\epsilon\ll\Gamma$ gets nearly total transmitted, in other word,
the NP part is the domination of the amplitude of the both photons transmitted, however, the IP part plays
the important role in the amplitude of the both photons reflected~\cite{JQLiaoPRA10,ShenOL05,ShenPRL05}.
\begin{figure}[tbp]
\includegraphics[width=8cm]{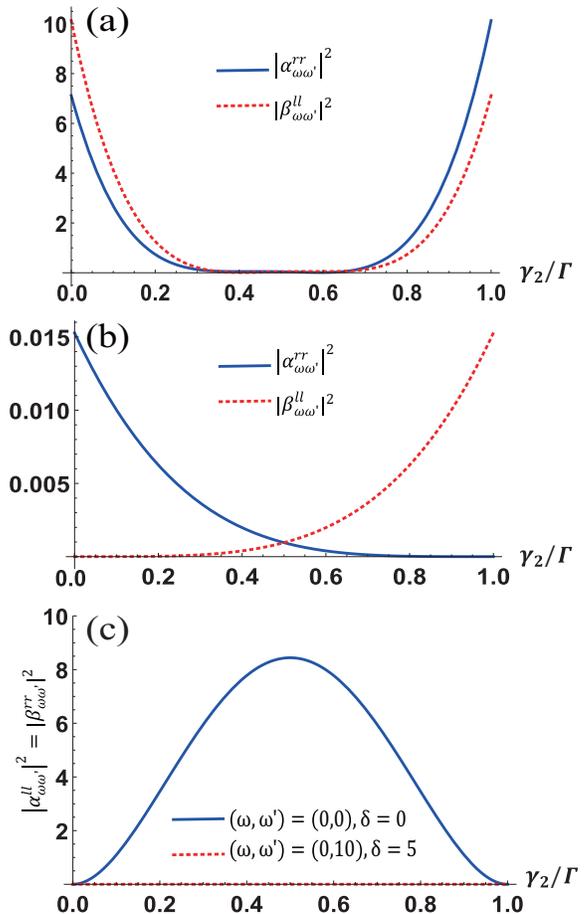}
\caption{(Color online) The norm square of the amplitudes of $\alpha_{\omega \omega'}^{rr}$ and $\beta_{\omega \omega'}^{ll}$
as functions of $\gamma_2$ for (a) $\omega=\omega'=\delta=0$, (b) $(\omega,\omega')=(0,10)$ or $(0,10), \delta=5$. (c) The probability
for the two reflected photons as functions of $\gamma_2$. The other parameters are set as follow:
$U=10, \epsilon=0.1$. All parameters are in units of $\Gamma$. }
\label{fig5.eps}
\end{figure}

To show the effect of the asymmetric coupling on the scattering wave of the two photons coming from the
left or right of the nonlinear resonator, we have plotted the norm square of the amplitudes of $\alpha_{\omega \omega'}^{rr}$
and $\beta_{\omega \omega'}^{ll}$ as functions of $\gamma_2$ by fixed $U=10, \epsilon=0.1$ at the single
photon resonance $\omega=\omega'=\delta=0$ and the two-photon resonance $(\omega,\omega')=(0,10)$ or
$(0,10), \delta=5$ as well as the two reflected photons as functions of $\gamma_2$ in Fig.\ref{fig5.eps}.
The asymmetric light transmission can be obtained as long as $\gamma_1\neq\gamma_2$. At the single photon resonance in
Fig.\ref{fig5.eps}(a), the probabilities for two transmitted photons first decreases as $\gamma_2$ increases
from 0 to $0.5$, then they increases as $\gamma_2$ increases, they only differ each other in magnitude,
we note that this phenomenon also appears when the position of the maximum value of the IP part is coincidence
with the center of the incident wave. At the two-photon resonance in Fig.\ref{fig5.eps}(b), the probabilities
for two transmitted photons $\alpha_{\omega \omega'}^{rr}$ incident from the left of the nonlinear resonator
decreases as $\gamma_2$ increases, but the probabilities for two transmitted photons $\beta_{\omega \omega'}^{ll}$
incident from the right side increases as $\gamma_2$ increases. The monotonicity in Fig.\ref{fig5.eps}(b)
is produced merely by the IP part because the incident photons are localized around $(\omega,\omega')=(5,5)$ under
the condition $\epsilon\ll \Gamma< \delta$. Since the asymmetric coupling permits unidirectional
propagation of light and forbids the transmission in the reverse direction, the nonlinear resonator
accomplishes the diode activity for two identical photons at the two-photon resonance with each energy
far away from the incident center.

%%%%%%%%%%%%%%%%%%%%%%%%%%%%%%%%%%%%%%%%%%%%%%%%%%%%%%%%%%%%

\section{\label{Sec:5}conclusion}

%%%%%%%%%%%%%%%%%%%%%%%%%%%%%%%%%%%%%%%%%%%%%%%%%%%%%%%%%%%%
In conclusion, we have studied the coherent transport of one or two photons in a 1D waveguide asymmetrically
coupled to a nonlinear resonator. We first construct the exact single-photon scattering states after
a long time, and find that the asymmetric coupling leads to a phase difference on the amplitude of
transmittance, a non-reciprocal phase shifter can be formed by the nonlinear resonator. Later, an analytic
solution of two-photon scattering is derived, it is found that 1) the two-photon scattering state incident
from either the left side or the right side of the nonlinear resonator contains the contribution of the NP
part and the IP part. The different phases of transmission amplitude caused by the chiral coupling produce
the different interference pattern of two-photon transmission at single-photon resonance. 2) The forward and reverse
transmitted probabilities change with the asymmetric coupling. The forward and reverse transmitted probabilities
have the same monotonicity as the asymmetric coupling changes and only differ each other in the magnitude as
the position of the maximum value of the IP part is coincidence with the center of the incident wave. However,
the forward and reverse transmitted probabilities have different monotonicity under the condition $\epsilon\ll \Gamma< \delta$
when the position of the maximum value of the IP part is far away from the center of the incident wave, so a
nonreciprocal two-photon transmission through a nonlinear resonator is found due to the energy redistribution
of the two photons by the bound state, which indicates that the nonlinear resonator acts as an optical diode.

\begin{acknowledgments}
This work was supported by NSFC Grants No. 11975095, No. 12075082, No.11935006,  No.12064010,
and the science and technology innovation Program of Hunan Province (Grant No. 2020RC4047)
\end{acknowledgments}

\end{document}